\documentclass[aps,prb,twocolumn,groupedaddress,showpacs]{revtex4}
\usepackage[english]{babel}

\usepackage{graphicx}

\usepackage{dcolumn}

\usepackage{bm}
\usepackage{epsf}
\usepackage{color}

\usepackage{color}
\usepackage{nth}

\begin{document}
\title{Molecular Dynamics and 
Object Kinetic Monte Carlo Study of Radiation Induced Motion of Voids and He Bubbles in BCC Iron} \author{G.J.~Galloway and G.J.~Ackland}
        
\affiliation{School of Physics, SUPA and CSEC, The University of Edinburgh, Edinburgh,
EH9 3JZ, UK. }

\date{\today}
\begin{abstract}

We show that voids adjacent to radiation damage cascades can be moved
in their entirety by several lattice spacings.  This is done using
molecular dynamics cascade simulations in iron at energies of 1-5~keV.
An equation describing this process is added to an an object kinetic Monte Carlo ({\sc okmc})
code to allow study of the mechanism at longer time scales.
The mechanism produces an enhancement of void diffusion by 2 orders of magnitude
from 1x10\textsuperscript{-22}~cm\textsuperscript{2}/s to
3x10\textsuperscript{-20}~cm\textsuperscript{2}/s.
Repeating the study on He bubbles shows that the movement is damped by the presence of
helium in the void.

\end{abstract}

\maketitle

\section{Introduction}

Steels under high energy neutron bombardment such as experienced in
fission and fusion reactors are known to develop voids and bubbles
over time.  These have seriously deleterious effects, and a clear
understanding of how these defects affect the properties of a material
is crucial to designing better steels.  Molecular dynamics can be used
to gain insight into the atomistic level interactions between
radiation cascades and defects such as voids and bubbles.  The long
term effects of the atomic mechanisms can be observed using object
kinetic Monte Carlo ({\sc okmc}) and such a multi-scale approach is
becoming increasingly popular.

A study of very low energy cascades of just a few eV impacting on
voids has been performed by Dubinko\cite{dubinko2005}.  It showed that
very low doses can lead to the dissolution of voids by focusons
striking them.  This is consistent with experimental
observation\cite{voidReductionExp} that the voids can be reduced in
size by irradiation under favourable conditions.

At higher energy Pu\cite{PuBubbleMD}~et~al. studied stability of
helium bubbles struck by radiation cascades.  They find that the
stability of the bubble depends on its He/vac ratio.  Clusters with
significantly more He than vacancies tend to absorb vacancies from the
cascade.  Conversely clusters with a low density of He tend to lose
vacancies during the cascade, approaching the stable ratio of near
1:1.

Simulations by Parfitt and Grimes\cite{grimes} study the behaviour of helium bubbles in uranium dioxide under radiation induced cascades.
An atom close to the bubble is given 10~keV of energy, to simulate a radiation event, and the resulting cascade observed.
They find that where the hot 'melt' region of the cascade overlaps the bubble, there is an increased chance of helium gas being emitted into the region.
As the melt region cools the helium is trapped in the lattice.
Their study concludes that this emission is mostly not ballistic,
but rather is due to the disorder of the bubble walls allowing low energy diffusion paths into the lattice.
Some ballistic emission of gas is also observed at shorter time scales.
Most of the gas remains confined to the bubble.
They expect the mechanism of increased surface diffusion to be dominant for larger bubbles.

Further properties of He in iron during cascades (but not in
the presence of voids or bubbles) have been obtained by
Lucas\cite{lucusHeinFe}~et~al.  They find that the presence of
substitutional He tends to reduce the number of Frenkel pairs
produced.  Interstitial He increases lattice stress and favours the
formation of self interstitial atoms (SIA) as well as stabilising
SIA clusters.  They observe increased diffusion of He at temperature,
which explains how bubbles can form rapidly.

An interesting mechanism for interstitial cluster formation during
cascades in pure $\alpha$-iron has been observed by
Calder\cite{bacon}~et~al.  They find that some cascades emit particles
moving faster then the cascade wave front, which cause secondary
cascades ahead of the main one.  Collision of the high density primary
wave with the core of the secondary
cascades can lead to nucleation sites for interstitial clusters.
This is due to atoms being forcefully pushed into the secondary
cascade's low density core region.

Under certain irradiation conditions voids can form
lattices\cite{voidOrdering}.  This effect has been reproduced by
Heinisch\cite{KMCvoidLattice}~et~al. in KMC simulations.  Their study
uses clusters of crowdions formed by irradiation and moving mostly one
dimensionally along close-packed planes to explain the alignment.
Voids off the lattice are exposed to a greater flux of crowdions and
so tend to shrink, whereas voids on-lattice are shielding each other
along the close packed planes.  The simulations do produce a void
lattice successfully although the lattice does not show the same
refinement as experimentally observed.

This current study looks at the interaction of voids and bubbles with
cascades and how this affects their centre of mass. 
Simulations show a mechanism  of atomic injection into the void by the density wave-front
of the cascade.  This leads to a new void being formed at the cascade
core, because the atoms absorbed by the original void are not
available for recombination.  The mechanism is explored atomistically in
molecular dynamics. 
A simple equation is fitted to model the motion observed in the molecular dynamics simulations.
This equation is then used as an event in the {\sc  okmc} simulations, along with other thermal processes,
 to allow the long term evolution of the system to be explored.

\section{Method}

The molecular dynamics study used a version of {\sc moldy}
\cite{moldy} modified for variable time step cascade simulations as
detailed previously\cite{CUARpaper}.  The Ackland-Mendelev\cite{AMS}
embedded atom potential was used for Fe-Fe interactions.  The Juslin
pair potential\cite{FeHeJuslin} was used for Fe-He and the
Beck\cite{HeBeck} pair potential for He-He.  The Beck He-He potential
was splined to the {\sc zbl}\cite{ZBL} universal potential at short range.
Helium is a closed-shell atom, and we do not expect any chemical bonding,
so the pair potential formalism is sensible.

The study consists of triggering radiation cascades in $\alpha$-iron
near to a void or bubble.  To create stable voids or bubbles in a 
{\sc bcc} iron lattice, a
single atom was removed near the centre of the system.  The system was
then relaxed and the atom with the highest potential energy removed.
Iterating this process produces a stable void with faceted sides
aligned to stable lattice directions.  Stability was tested by running
the sample at 900~K for 20~ps and observing no changes.

Bubbles were produced by placing helium atoms onto the vacant sites
in the void.  The system was then quenched using low temperature MD
with a high thermostat value to limit the velocity of the helium
atoms.  Once the helium has reached a stable configuration the system
may be thermally equilibrated with standard thermostating.

Cascades were initiated by giving a substantial kinetic energy to one
iron atom near to the defect cluster, this atom is called the primary
knock-on atom ({\sc pka}).  Cascade energies of 1, 2.5 and 5~keV were
considered.  For 1 and 2.5~keV cascades a box dimension of $40~a_0^3$
(128000 atoms) was used, with $65~a_0^3$ (549250 atoms) for 5~keV.  We
use the NVT canonical ensemble (constant number, volume, cell shape,
temperature and zero total momentum) 
with periodic boundaries.

The final atomic configurations were analysed by comparison to a
perfect lattice aligned to the defective lattice\cite{CUARpaper}.  The
Wigner Seitz cell around each perfect lattice point is checked for
atoms, allowing easy identification of interstitials, vacancies and
substitutional defects.  Defects were considered to be clustered if
they were within 2$^{nd}$ nearest neighbour distance of each other.

The data was further processed by comparing the nearest neighbours
of every atom before and after the collision, marking atoms with at
least 2 new neighbours.  This identifies the area that has been
affected by the cascade, as well as being sensitive to correlated
motion of atoms.  Additionally it provides an effective method for
visualising the cascade by joining a line from the initial to final
location of each atom (c.f. Bacon\cite{bacon}~et~al.).  The region
covered by these lines is the part of the simulation ``melted'' by the
cascade and the lines tend to form connected paths due to
replacement-collision events.  By taking the list of atoms that have
changed neighbour, and filtering it so that only members that have
moved at least one lattice site remain, an estimate of cascade volume
is obtained.  The cascade radius is calculated from this volume.

The point around which the cascade is centred is identified by the
centre of mass of those listed atoms with at least four neighbours
also in the list.  This eliminates small sub-clusters and
replacement-collision chains that would otherwise skew the estimate.

\begin{figure}     
\begin{center}
\includegraphics[width=80mm]{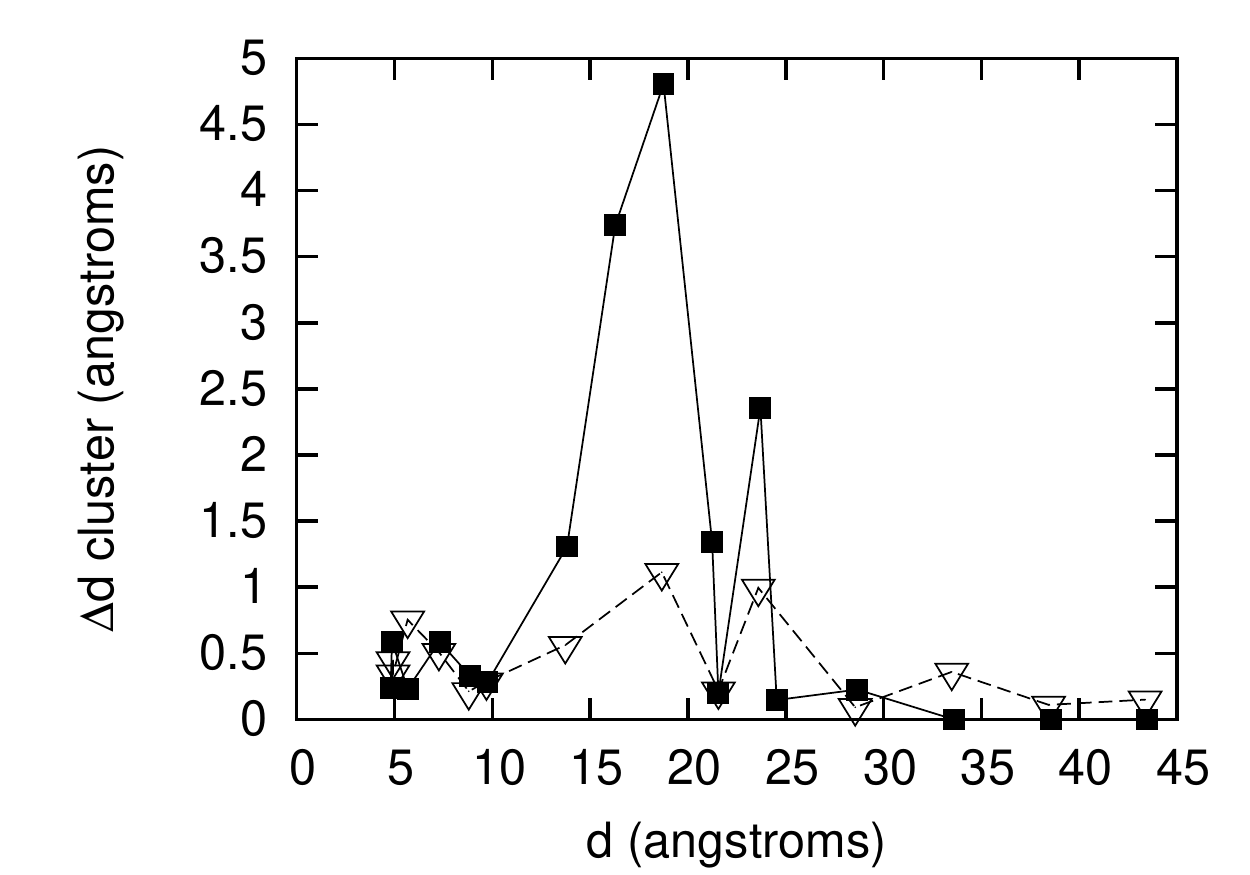}
\end{center}
\caption{Graph showing movement of void centre of mass from starting position, $\Delta d\, {\rm cluster}$, (filled squares) and bubble (empty triangles) vs. distance to {\sc pka}, d, for a 30 defect cluster hit at 1~keV.
It can be seen that the void can move significantly towards the {\sc pka} site for certain cascade distances.
The presence of helium in the void greatly damps this effect.
\label{fig:1keVMovement}}
\end{figure}

\begin{figure}     
\begin{center}
\includegraphics[width=80mm]{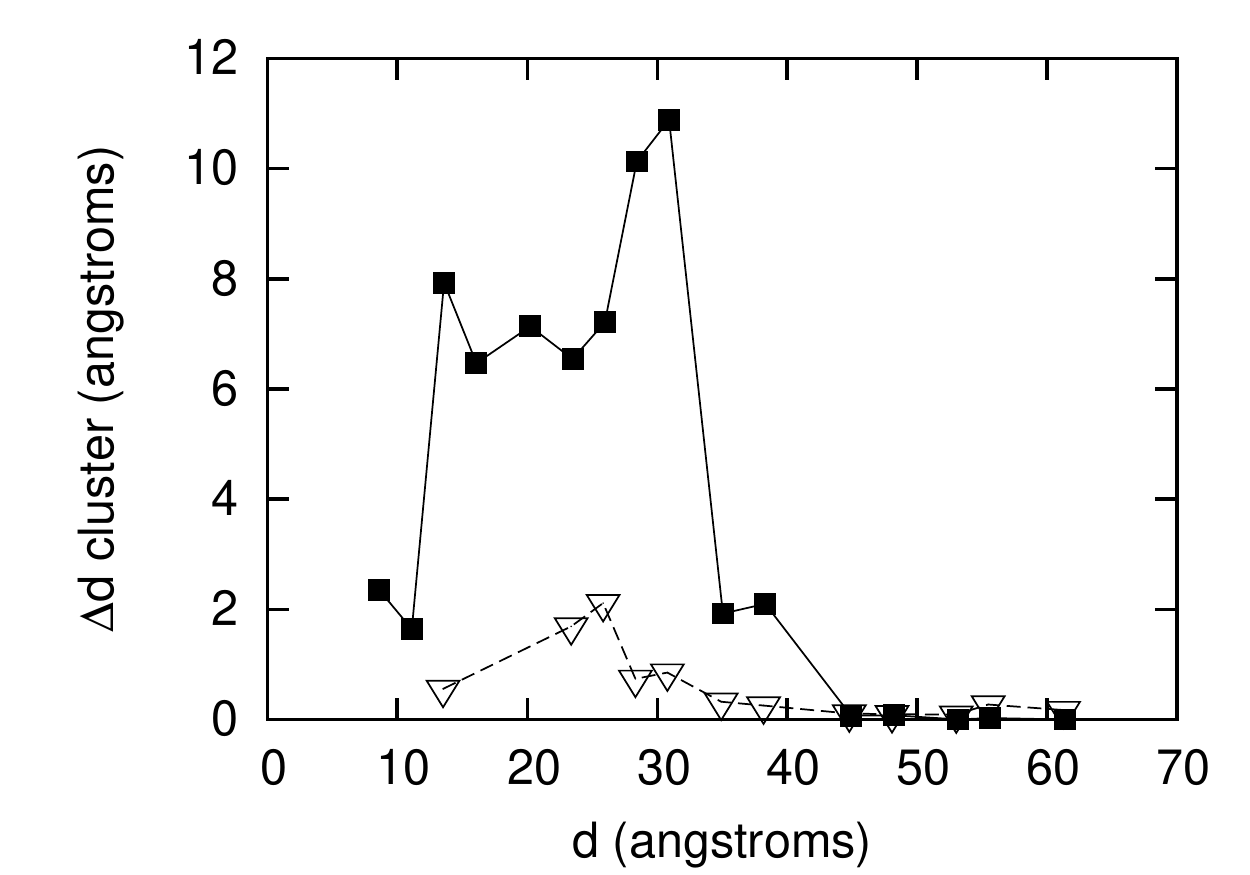}
\end{center}
\caption{Graph showing movement of void centre of mass from starting position,  $\Delta d\, {\rm cluster}$, (filled squares) and bubble (empty triangles) vs. distance to {\sc pka}, d, for a 100 defect cluster hit at 5~keV.
It can be seen that the void can move significantly towards the {\sc pka} site for certain cascade distances.
The presence of helium in the void greatly damps this effect.
The effect is much more pronounced than at 1~keV.
\label{fig:5keVMovement}}
\end{figure}

\begin{figure}     
\begin{center}
\includegraphics[width=80mm]{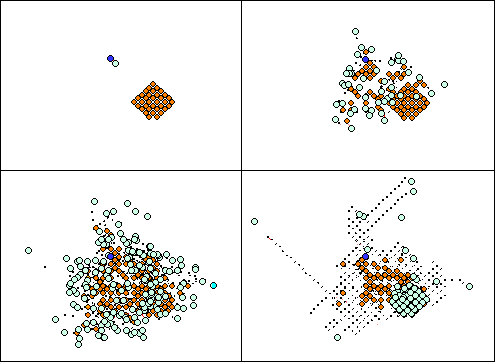}
\end{center}
\caption{(Colour online) The mechanism for cascade-induced void
  movement is shown above for a 5 keV collision with a 100 vacancy
  void.  The blue (dark grey) dot represents the starting positions of the
  cascades, orange dots (grey) are vacancies, light blue dots (white) are
  interstitials or filled vacancies.  Small black dots represent the
  atoms that have moved in the cascade.  Non-defective atoms are
  omitted and the full box size is not shown.  Top left: Void and {\sc
    pka} positions and {\sc pka} direction shown after 2~fs.  Top
  right: Wave of displaced atoms starts to form and expand at 76~fs.
  Bottom left: Full extent of wave, encompasses the void, injecting
  atoms into its core at 260~fs.  Bottom right: Wave collapses back,
  but atoms are trapped in void, leaving a
  depletion at cascade core by around 1~ps (image taken at 5~ps for
  clarity).  Note if an initially-vacant site in the void is filled,
  it is coloured light blue (white).  This was done to show the filling of
  the void.
\label{fig:formation}}
\end{figure}

\begin{figure}     
\begin{center}
\includegraphics[width=80mm]{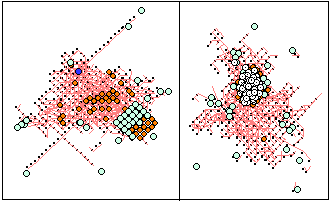}
\end{center}
\caption{(Colour online) a) (left): Cascade radius overlaps half the void leading to partial filling and splitting of the void. 
Half the void is unaffected, and now forms a smaller void.
A second void is formed at the depleted cascade core, as several atoms from the cascade are now trapped in the near side of the old void, preventing their return.
The shaded region represents atoms that have been displaced by the cascade, and is related to the cascade radius.
 Orange dots (grey) are vacancies, light blue dots (white) are interstitials  or filled vacancies.
b) (right): Typical final state of a 5~keV cascade impacting 100 He bubble in a 100 vacancy void.
He remains clustered in the void. He atoms are white with a dot.
\label{fig:splitVoidBub}}
\end{figure}

\section{results}

Simulations of 
a cascade interacting with a void or bubble 
in a
periodic cubic supercell of $\alpha$-Fe were performed under various
conditions.  Most pkas were given initial velocities close to
$\langle111\rangle$ direction towards the void, with the slight
deviation to avoid perfect replacement-collision chains.  This high
symmetry direction was chosen in order to localise the cascade\cite{nochannel} and
give easily-comparable data across different simulations.

To explore if the effects observed are dependent on the crystal
alignment some cascades were also performed in the $\langle311\rangle$
direction and in random directions.  Simulations were performed at 1,
2.5 and 5~keV on voids/bubbles of size 30 and 100, with the bubbles at
ratio 1:1 for vacancy:helium.  One set of data is also performed at
300~K to explore temperature effects (the various simulations along
with their damage yields are summarised the supplementary material).

For 300~K simulations little change was seen in the results apart from
a slight reduction of Frenkel pairs, and so further temperature
studies were not pursued, although higher temperatures may have more
effect.  Results along the $\langle311\rangle$ direction were found to
be consistent with similar studies of the $\langle111\rangle$
direction.  The radii and damage of the cascades was also consistent
across randomly launched trajectories.

Normally, voids move by capture and emission of point defect, but
interestingly some of our simulations show that the void is moved by
several angstroms, far more than absorption of the defects created by
the cascade would allow.
The process appears to have an upper and lower
interaction radius, close to the size of the cascade, as shown in
Fig.~\ref{fig:1keVMovement}~and~\ref{fig:5keVMovement}.

A typical cascade displaces atoms around itself creating a density
``wave'' moving spherically outwards, with a depleted region at its
core.  Normally in a perfect lattice this ``wave'' will collapse back
as the energy dissipates, causing recombination of vacancies and
interstitials at the core (Figure \ref{fig:formation}).  The defect
yield ({\sc dpa}) from a cascade normally considers only point defects
surviving this recombination.

However, when a pre-existing void is adjacent to the cascade a more
complex mechanism may take place.  The density ``wave'' injects atoms
into the pre-existing void, where they find sites to occupy and so do
not return as the wave collapses back.  This results in the original
void being filled, and a new void of equal size appearing at the
cascade core.  The outcome of this process is equivalent to the
movement of the entire void by several lattice spacings.  This
behaviour was reproducible for voids under all conditions considered,
providing the cascade overlaps the void.

Figures~\ref{fig:1keVMovement}~and~\ref{fig:5keVMovement} show how the
distance moved by the centre of the void
depends on the separation between void and cascade.
The
cascade will typically begin a few lattice spacings from the starting
position of the {\sc pka}.  At very short distances the {\sc pka} can
pass directly through the void, which is why there is a minimum
interaction threshold for {\sc pka} distance.  Data suggests
interaction thresholds of 12~\AA~ and 25~\AA~ with peak interactions
around 18~\AA~ for 1~keV (Fig.~\ref{fig:1keVMovement}).  For 2.5 keV
the interaction starts at 18~\AA~, lasting until 40~\AA~ and has peak at
31~\AA.  For 5~keV the interaction starts around 12~\AA~ ends at
38~\AA~ and had peak interaction at 31~\AA~
(Fig.~\ref{fig:5keVMovement}).

For 1keV cascades on bubbles of size 30 almost no motion of the
cluster is observed even for high cascade overlap.  The slight
movement seen (Fig.~\ref{fig:1keVMovement}) is due to the side of the
bubble nearest the cascade absorbing a few vacancies, shifting the
centre of the cluster.  On average the bubble grew by one vacancy and
almost all the helium remained clustered in it.

For a void of size 100 impacted at 5~keV, significant movement of the
void towards the cascade was observed (Fig.~\ref{fig:5keVMovement}).
Void size tended to decrease slightly on average to 84 vacancies.
This change is largely due to the void being split,
 as shown in Fig.~\ref{fig:splitVoidBub}a, which occurs
in 3 simulations.  Ignoring the split samples void size tends to be
around 88 after cascade interaction.

Similar results to the 1~keV bubble case were observed for a 5~keV cascade impacting a helium bubble
of size 100.  Fig.~\ref{fig:splitVoidBub}b shows a typical bubble
after impact.  The helium reduces the mobility of the cluster
(Fig.~\ref{fig:5keVMovement}) and the helium remains clustered in the
bubble.  Cascade radius is unaffected, although the number of Frenkel
pairs formed seems to increase slightly (table~\ref{table:yields}).
Occasional shrinkage by up to 5 vacancies is observed at longer
distance due to interstitial absorption from the cascade periphery.

The standard deviation of the vacancies in the voids and bubbles were analysed to explore the compactness of the cluster.
For each void, the standard deviation of the vacancies in the void is calculated, and then this number is averaged over different voids.
The standard deviation for clusters that move more than 1~\AA~ during the cascade is compared to the value for a perfect cluster, see table~\ref{table:stdev}.

Results shows that the bubbles remain compacted, while the voids become more diffuse.
Since all the defects are within 2$^{nd}$ nearest neighbour distance of each other in a cluster,
it is expected that they will collapse to a faceted void due to vacancy attraction.
In kinetic Monte Carlo models it is standard to assume that there is a capture radius of at least this distance\cite{Caturla,Caturla2,okmcDomain,okmcSoneda}
 between vacancies, leading to collapse to a compact void.

\begin{table}[ht]
\caption{The standard deviation of vacancies in the clusters: $ \sigma = (1/(N-1)) \sum_{i=1}^N (r_i-<r_o>)^2  $
where $r_i$ are the location of the vacancies in the void.
the standard deviation of vacancies for each cluster is calculated as a measure of diffuseness of the cluster.
Size refers to the number of vacancies in the cluster. Perfect~$\sigma$ is the standard deviation for a perfect cluster.
bubble~$\sigma$ and void~$\sigma$ refer to the average standard deviation for bubbles and voids after cascade impact.
Only clusters that move at least 1~\AA~ are included. } 
\centering 
\begin{tabular}{c c c c } 
\hline\hline 
\\ [-1.5ex]
size & ~perfect~$\sigma$~(\AA)~~~ & ~~bubble~$\sigma$~(\AA)~~~ &  ~~void~$\sigma$~(\AA)~~~   \\ [0.5ex] 
\hline  \\ [-1.5ex]
30  & 3.4 & 3.7 & 4.8\\
100 & 5.1 & 5.4 & 6.6\\

\\ [-1.5ex]
 
\hline 
\end{tabular}
\label{table:stdev} 
\end{table}

\begin{table}[ht]
\caption{The Frenkel pair count, $N_f$, and cascade radius, $r_c$, for various simulations. E is the {\sc pka} energy. } 
\centering 
\begin{tabular}{c c c c} 
\hline\hline 
\\ [-1.5ex]
~~~E~(keV)~~~ & ~~~~~$r_c$~(\AA)~~~~~ & type & ~~~~~$N_f$~~~~~   \\ [0.5ex] 
\hline  \\ [-1.5ex]
1  & 7.3 & void &5.5\\
1&7.2&bubble&6.2\\
2.5  & 10.8 & void&11.2 \\
5 &   15.1 &void&16.7\\
5& 15.3 & bubble &19.1\\
\\ [-1.5ex]
 
\hline 
\end{tabular}
\label{table:yields} 
\end{table}

\begin{table}[ht]
\caption{Model parameters: Table of migration energy, $E_m$, and dissociation energy of point defects from clusters, $E_d$.
Formation energies: $E_f^V = 2.07$ and $E_f^I = 3.77$~eV.
$E_b$ is the binding energy of a cluster with two defects, $E_b^V=0.30$ and $E_b^I = 0.80$~eV. All diffusion is 3D.
$g(n) = \frac{n^{2/3}-(n-1)^{2/3}}{2^{2/3}-1}$} 
\centering 
\begin{tabular}{c c c c c} 
\hline\hline 
\\ [-1.5ex]
~~~Type~~~ & ~~~~~$E_m$~(eV)~~~~~  &  ~~~~~$E_d$~(eV)~~~~~   \\ [0.5ex] 
\hline  \\ [-1.5ex]
$V$& 0.67  & \\
$V_2$ & 0.62 & 0.97\\
$V_3$ & 0.35  & 1.04\\
$V_4$ & 0.48   & 1.29\\
$V_n,n>4$ & immobile   & $E_m^{n=1}+E_f^V+(E_b^V-E_f^V)g(n)$\\
$I$ & 0.34   & \\
$I_2$& 0.42 & 1.14\\
$I_3$ & 0.43   & 1.26\\
$I_4$ & 0.43  & 1.26\\
$I_n,n>4$ & immobile   & $E_m^{n=1}+E_f^I+(E_b^I-E_f^I)g(n)$\\
 [1ex] 
\hline 
\end{tabular}
\label{table:OKMC} 
\end{table}

\section{OKMC rates}

In this work we wrote an Object kinetic Monte Carlo ({\sc okmc}) code to study how voids repeatedly exposed to radiation cascades move,
on a timescale far longer than could be achieved in molecular dynamics. 
In addition to thermally-induced vacancy and interstitial cluster motion,
we include defect creation from cascade debris, and allow vacancy clusters to jump to cascade sites. 
These latter two rates are dose-dependent,
and their effect is modelled using data gained from the molecular dynamics simulations.

Object kinetic Monte Carlo ({\sc okmc}) is similar to normal kinetic Monte Carlo ({\sc kmc}) of defects, but allows clusters of defects to be treated as objects.
Clusters can have their own individual diffusion rates, dissociation rates and interaction events that affect the cluster as a whole,
instead of dealing with the constituent point defects.
This method is more efficient than {\sc kmc} as the number of bodies in the simulation is greatly reduced.
Other methods such as accelerated molecular dynamics\cite{AMD} also allow the time scale to be increased beyond molecular dynamics timescales,
and have the advantage of not requiring pre-defined rate tables. 
However, given the time scales that we wished to explore and the availability of data for rates,
including input gained from the molecular dynamics simulations, {\sc okmc} was chosen.

The {\sc okmc} code was written and used to explore the significance of long term evolution of the void motion.
The short time-scale jump events identified in the molecular dynamics are added as an equation in the {\sc okmc} and are not directly atomistically simulated.
Jump events are simulated by moving the void to the cascade core if it is within the interaction radius of the cascade.
These jumps can then be simulated alongside other mechanisms such as radiation induced Frenkel pairs and thermal dissociation.
Locally the motion induced by the cascade is ballistic in nature, as the void is essentially moved by the expanding wave of interstitials.
However, when considered over longer time scales the motion is diffusive in that the void makes jump at a given rate, undergoing a random walk.
The diffusion rate for such a process can be calculated as $D = x^2 / (6t) $,
where x is the average jump distance and t is the average time between jumps.
This is not thermally activated diffusion, rather it is activated by proximity to a cascade.

The code implements the standard resident-time algorithm, also
called the kinetic Monte Carlo or Gillespie
algorithm\cite{Gillespie1976}.  The established mechanism for void
diffusion is by point defect absorption and emission events (later
referred to as standard diffusion) and so our new mechanism's
diffusion is compared to this process.  The OKMC model includes
vacancy clusters, interstitial clusters, defect creation from cascade
debris and direct cascade-void interactions.  It allows diffusion of
small clusters and thermal dissociation of point defects from clusters
following the energy barriers given by the
Fu\cite{FuDFT}~et~al. study as used in
Caturla's\cite{caturlaPico}~et~al. model (see table
\ref{table:OKMC}).  The capture radii of clusters follows the model of
Caturla\cite{Caturla}~et~al.

\[ r_{I_n,V_n} = Z_{I/V}\left(\left(\frac{3n\Omega}{4\pi}\right)^{1/3}+r_0\right) \]

where $r_{I_n,V_n}$ are the capture radius for interstitial and vacancy clusters respectively.
$Z_I = 1.15$ is the bias factor accounting for the increased strain field observed for interstitials.
$\Omega$ is the atomic volume and $r_0 = 3.3$~\AA.

Open absorbing boundaries were used for this study, simulating a single
void at the centre of a single grain.
Radiation cascades were introduced into the sample with the maximum interaction radius taken from the data sets in the current study (Figs.~\ref{fig:1keVMovement}~and~\ref{fig:5keVMovement}).
The cascade debris is calculated by the Bacon\cite{baconNvsE}~et~al. empirical formula for iron: $N_f = 5.67E^{0.779}$, where $N_f$ is the number of Frenkel pairs and E is the damage energy in keV.
Cascades were implemented in a shell like manner, with vacancies randomly placed in a smaller radius, and surrounded by a shell of interstitials.
Cluster interactions were carefully performed to take into account the correct change in cluster position for absorption and emission events, as this will affect the diffusion.

If the cascade is sufficiently close to the void, the void is shifted to the centre of
the cascade region as seen in the MD simulations.  By checking the
interaction volume of the cascade for the largest void within distance
and updating according to:

\[
{\rm if}~r_{cascade}+ r_{void}> d_{casc-void}~{\rm then}~\vec{r}_{void}=\vec{r}_{cascade}\]

 $r_{cascade}$ is the cascade radius, $r_{void}$ the void radius, $d_{casc-void}$ is the distance from the cascade centre to the void centre, and $\vec{r}_{void}$/$\vec{r}_{cascade}$ is the position vector for the void/cascade.
This mechanism was disabled to simulate the standard diffusion.

The trend of both diffusion mechanisms with temperature, dose rate,
cascade energy and system size was studied.  The contribution from
each process was calculated by separating out the jumps due to each
mechanism.  Studies were performed with default values: Temperature of
500~K, dpa/s of 1x10\textsuperscript{-6}, cubic cell of 100~\AA~ and
cascade energy of 10~keV on a void of 100 vacancies.  Only one
parameter was varied from default at a time, to gain information on
system trends without performing an exhaustive search of the parameter
space.  Parameter ranges considered: Temperature of 400-600~K, dpa/s
of 10\textsuperscript{-8}-10\textsuperscript{-2}, cubic cell of
50-200~\AA~ and cascade energy of 1-20~keV.  Under the default
parameters the standard diffusion was found to be
1x10\textsuperscript{-22}~cm\textsuperscript{2}/s while the diffusion
due to the new mechanism is
3x10\textsuperscript{-20}cm\textsuperscript{2}/s , two orders of
magnitude larger.

For fixed dpa rate, the standard diffusion rate is found to not be
strongly dependent on cascade energy.  The mechanism only depends on
the number of free point defects and increasing cascade energy
decreases the number of cascades to conserve dpa rate.  however, the new
mechanism is highly dependent on cascade energy, because
this sets the interaction radius.

System size does not strongly affect the new mechanism.
It does affect the standard mechanism as the only initial sinks in the system are the void and cell walls.
Increasing cell size gives defects more chance to be absorbed by the void, instead of the walls and so effectively increases the void capture radius.
This leads to a slight increase in diffusivity.

Rates for both mechanisms increase approximately
linearly with dose rate so this parameter is not directly critical,
although it is important in setting the relative time scale of the
system as compared to dissociation events.

Neither mechanism is strongly temperature dependent. 
The standard mechanism lacks temperature dependency because the diffusion of point defects is much faster than the time between cascades.
Temperature will only become critical at lower T when diffusion time of point defects is comparable to the time between cascades.

The new mechanism is found to be dominant over the standard mechanism at all parameters considered apart from for energies of 2~keV or below, when the interaction radius becomes too small to have significant effect.
Cascade energy is the most significant factor, with diffusion of the new mechanism rising to 1x10\textsuperscript{-19}~cm\textsuperscript{2}/s at 20~keV.

\section{Conclusions}

We have shown that the migration rates of voids in irradiated
environments are much higher than expected from simple point defect
migrations.  Cascades occurring adjacent to voids may cause a jump of
the entire void by several lattice spacings, if the cascade is within
appropriate interaction thresholds.  This motion is due to the
injection of atoms by the cascade into the original void, trapping
them and leaving a new void at the cascade core.  The effect becomes
more significant as cascade energy increases (even at fixed dpa)
because it depends on cascade radius.  This jump mechanism has
been shown in {\sc okmc} simulations to be a more significant mechanism than
diffusion due to absorption and emission of point-defects.
The mechanism is suppressed when helium is present in the void.

The rate is equivalent to one 10.5~\AA~ jump every 18.6 hours on average.
This is still a much faster process than the standard diffusion mechanism.
Although the overall void diffusion rate is low, the mechanism may still play a role in void mobility.
Jumps displace the entire void by several lattice spacings.
This may have impact on long-term void evolution processes, such as void lattice formation.
Assuming the alignment method discussed in the introduction, involving $\langle111\rangle$ interstitial 
clusters\cite{KMCvoidLattice},
jumps may expose or shield voids from this flux of interstitials. 
These occasional jumps may help to fine-tune the void lattice over long periods of time.
Further study would be needed to verify this.

\section*{Acknowledgements}

We thank D.~J.~Hepburn for useful discussions and the EPSRC Scottish CM-DTC for a studentship (GJG).

\bibliographystyle{unsrt}
\bibliography{ref}

\begin{thebibliography}{10}
\bibitem{nochannel}
For some initial directions the pka can be channelled between atomic layers,
  causing the cascade to occur far from the void.
\end{thebibliography}


\newpage
\begin{widetext}
\appendix
\section{Supplementary: Simulation data}

\begin{table*}[h-]
\caption{Supplementary table of simulations: Voids.
 $N_f$ is the number of Frenkel pairs formed.
 $V_{void}$ is the number of vacancies in the largest cluster.
 $r_c$ is the radius of the cascade.} 
\centering 
\begin{tabular}{c c c c c c c} 
\hline\hline 
\\ [-1.5ex]
~~~E~(keV)~~~ & ~~~setup~~~ & simulations &defect-size& ~~~~$N_f$~~~~&   $V_{void}$ &~~~~~$r_c$~(\AA)~~~~~     \\ [0.5ex] 
\hline  \\ [-1.5ex]
1 & $\langle111\rangle$ & 17 & 30 & 5.3 & 28.8 & 7.1\\
1 & random & 48 & 30 & 5.8 & 29.8 & 7.2\\
1 & $\langle311\rangle$  & 10 & 30 & 5.9 & 29.6 & 7.5\\
1 & $\langle111\rangle$, T=300~K & 14 & 30 & 4.6 & 28.1 & 7.6\\

2.5 & $\langle111\rangle$ & 12 & 30 & 14.2 & 30.8 & 11.0\\
2.5 & random & 16 & 30 & 9.9 & 26.8 & 10.6\\
2.5 & $\langle111\rangle$ & 9 & 100 & 9.4 & 89.6 & 10.9\\

5 & $\langle111\rangle$ & 12 & 30 & 18.8 & 27.9 & 15.2 \\
5 & $\langle111\rangle$ & 16 & 100 & 14.6 & 84.3 & 15.1 \\


\\ [-1.5ex]
 
\hline 
\end{tabular}
\label{table:SuppVoid} 
\end{table*}

\begin{table*}[h-]
\caption{Supplementary table of simulations: Bubbles.
 $N_f$ is the number of Frenkel pairs formed.
 He$_{bub}$ is the helium remaining in the bubble after the cascade.
 He$_{sub}$ is the helium that has become substitutional in the lattice.
 He$_{int}$ is the helium that has become interstitial in the lattice.
 $r_c$ is the radius of the cascade.
 V$_{bub}$ is the number of vacancies in the bubble.
 He$_{bub}$ is the number of helium atoms in the bubble.
} 
\centering 
\begin{tabular}{c c c c c c c c c c} 
\hline\hline 
\\ [-1.5ex]
~~~E~(keV)~~~ & ~~~set~up~~~& simulations & He:vac bubble &  ~~~$N_f$~~~ & V$_{bub}$ &  He$_{bub}$ & He$_{sub}$ & He$_{int}$ & ~~~~~$r_c$~(\AA)~~~~~ \\ [0.5ex] 
\hline  \\ [-1.5ex]
1 & $\langle111\rangle$ & 14 & 29:30 & 6.5 & 31.1 & 28.3 & 0.6 & 0.1 & 7.4\\
1 & random & 46 & 29:30 & 6.1 & 30.2 & 28.9& 0.1 & 0 & 7.1\\
5 &  $\langle111\rangle$ & 12 & 100:100 & 19.1 & 102.2 & 100 & 0 & 0 & 15.3\\

\\ [-1.5ex]
 
\hline 
\end{tabular}
\label{table:SuppBubble} 
\end{table*}

\end{widetext}

\end{document}